\documentstyle[11pt]{article}

\begin{document}

\title{STANDARD SOLAR NEUTRINOS}
\author{Arnon Dar \\
  Department of Physics \&
  Asher Space Research Institute \\
  Technion-Israel Institute of Technology,
  Haifa 32000, Israel}

\maketitle

\begin{abstract}
The predictions of an improved standard solar model are compared
with the observations of the four solar neutrino experiments. The
improved  model includes premain sequence evolution, element diffusion,
partial ionization effects, and all the possible nuclear reactions
between the main elements. It uses updated values for the initial
solar element abundances, the solar age, the solar luminosity,
the nuclear reaction rates and the radiative opacities.
Neither nuclear equilibrium, nor complete ionization are assumed.
The calculated $^8$B solar neutrino flux is consistent, within the
theoretical and experimental uncertainties, with the solar neutrino
flux measured by Kamiokande. The results from the $^{37}$Cl and
$^{71}$Ga radiochemical experiments seem to suggest that the terrestrial
$^7$Be solar neutrino flux is much smaller than that predicted. However,
the present terrestrial ``defecit'' of $^7$Be solar neutrinos may be due
to the use of inaccurate theoretical neutrino absorption cross sections
near threshold for extracting solar neutrino fluxes from production
rates. Conclusive evidence for a real deficit of $^7$Be solar neutrinos
will require experiments such as BOREXINO or HELLAZ. A real defecit of
$^7$Be solar neutrinos can be due to either astrophysical reasons or
neutrino properties beyond the standard electroweak model.
Only future neutrino experiments, such as SNO, Superkamiokande,
BOREXINO and HELLAZ, will be able to provide conclusive evidence that
the solar neutrino problem is a consequence of neutrino properties
beyond the standard electroweak model. Earlier indications may
be provided by long baseline neutrino oscillation experiments.

\end{abstract}

\section{ Introduction}
The sun is a typical main sequence star which
generates its energy by fusion of protons into Helium nuclei
through the pp and CNO nuclear reactions chains which
also produce neutrinos. These neutrinos have been detected
on Earth in four pioneering solar neutrino experiments, the
radiochemical Chlorine experiment at Homestake (Cleveland et al. 1995
and references therein), the electronic light water Cerenkov
experiment at Kamioka (Totsuka, these proceedings) and
the two radiochemical Gallium experiments, GALLEX at Gran Sasso
(Kirsten, these proceedings) and SAGE at the Baksan
(Gavrin, these proceedings). They provide
the most direct evidence that the sun generates its energy via
fusion of Hydrogen into Helium. However, it has been claimed
(e.g., Bahcall 1995, Hata et al. 1994) that all four experiments
measured solar neutrino fluxes significantly smaller than those
predicted by standard solar models (SSM) (e.g., Bahcall and Ulrich
1988; Turck Chieze et al. 1988; Sackman et al. 1990, Bahcall and
Pinsonneault 1992 (BP92), Turck-Chieze and Lopes 1993 (TL93); Castellani
et al. 1994; Kovetz and Shaviv 1994; Christensen - Dalsgaard 1994;
Shi et al. 1994; Bahcall and Pinsonneault 1995 (BP95);
See, however, Dar and Shaviv 1994 (DS94); Shaviv (1995): Dzitko et al.
1995; Dar and Shaviv 1995 (DS95).

This discrepancy has become known as the solar neutrino problem.
Three types of solutions to the solar neutrino problem have been
investigated:

(a) Experimental Solution:
Perhaps the accuracy of the results of the
solar neutrino experiments has been
overestimated and unknown systematic errors
are largely responsible for the solar neutrino problem.

(b) Astrophysical Solutions:
Perhaps the standard solar models do not provide sufficiently
accurate description of the present day sun and/or the neutrino producing
reactions in the sun.

(c) New Physics Solutions:
Perhaps new electroweak physics beyond the standard electroweak model
is responsible for the solar neutrino problem.

The chances that possibility (a) is responsible for the solar
neutrino problem have been greatly reduced by the GALLEX Chromium
source experiment performed last year (Anselmann et al. 1995).
This experiment is the first full demonstration of the reliability
of the radiochemical technique for the detection of solar neutrinos.
In particular, it excludes the possibility of any unidentified important
sources of systematical errors, such as hot atom chemistry, in
the radiochemical experiments.

Astrophysical solutions have been ruled out by many authors on the
grounds
that the combined results from the four solar neutrino experiments
show a large deficit of $^7$Be solar neutrinos relative to that
expected from the $^8$B solar neutrino observations in Kamiokande
(the ratio between proton capture in the sun by $^7$Be, which produces
$^8$B neutrinos, and electron capture by $^7$Be, which produces $^7$Be
neutrinos, is almost model independent):

1) The Cl experiment with an energy threshold lower than that of
Kamiokande measured a neutrino capture rate smaller than that expected
from the $^8$B solar neutrino flux measured
in Kamiokande, leaving no room for a significant $^7$Be
solar neutrino flux.

(2) The Gallium experiments measured
a neutrino capture rate which is consistent with
the solar luminosity only if the sun does not produce a significant
flux of $^7$Be  neutrinos.

Consequently, it has been concluded (see e.g., Bahcall
these proceedings) that non standard neutrino properties,
such as neutrino flavour mixing, are
required in order to explain the solar neutrino observations.

However, I do not consider yet the $^7$Be solar neutrino problem as a
solid evidence for new electroweak physics,
as claimed by various authors (e.g., Bahcall and Bethe 1993;
Bludman et al. 1993; Castellani et al. 1994; Hata et al. 1994;
Berezinsky 1994; Kwong and Rosen 1994;
Bahcall 1994; Parke 1995; Hata and Langacker 1995). This is because the
standard solar models are only approximate and simplified
descriptions of the real and complex sun, because of very little
experimental and theoretical knowledge of dense plasma
effects on nuclear reaction rates and on energy transport at
solar conditions, and because standard physics solutions are
still possible. In particular, below I will argue that:

i) The $^8$B solar neutrino flux observed by Kamiokande
is consistent with the predictions of the standard solar models
of Dar and Shaviv (1995) and of Dzitko et al. (1995).

ii) Suppression of the neutrino capture rates near threshold
in $^{37}$Cl and in $^{71}$Ga may be responsible for
the discrepancy between the observed capture rates of solar
neutrinos in $^{37}$Cl and in $^{71}$Ga and those calculated
from the standard solar model.

\section{ The Standard Solar Model - An Overview}
The standard solar model (e.g. Bahcall 1989 and references therein)
is a physical description of the sun based on
the standard stellar evolution equations (e.g., Clayton 1968),
which are used to calculate its evolution from its
premain sequence Hayashi phase to its present state.
The model assumes a complete spherical symmetry, no mass loss or
mass accretion, no angular momentum gain or loss, no differential
rotation and a zero magnetic field through the whole solar evolution .
It uses best available input physics (
equations of state, nuclear cross sections, radiative opacities,
condensed matter effects) and the following initial conditions:

\begin{enumerate}

\item  {Fully convective, homogeneous, spherically symmetric
 protostar.}

\item {Initial mass of $M_\odot=1.99\times 10^{33}gm$.}

\item { Zero angular momentum, no differential rotation, no magnetic
field.}

\item {Initial chemical composition deduced from primitive meteorites,
the solar photosphere, the solar wind, the lunar soil, the interstellar
medium and the photospheres of nearby stars.}
\end{enumerate}

The calculations are iterated, treating the unknown initial $^4$He
abundance and the mixing length in the convective zone (roughly the size
of the pressure scale height) as adjustable parameters, until the
present observed properties of the sun are reproduced at its
present estimated age, $t_\odot=4.57~GY.$
These include (Particle Data Group 1994):

\begin{enumerate}

\item {The solar luminosity $L_\odot=3.844\times 10^{33} erg\cdot
sec^{-1}.$}

\item { The solar radius $R_\odot=6.9599\times 10^{10}cm.$}

\item { The observed solar surface element abundances.}

\item { Internal structure consistent with helioseismology.}

\end{enumerate}

The output of the calculations includes the present-day density
profile $\rho(r)$, temperature profile $T(r)$ and chemical composition
profile $[X_i(r)]$ of the sun. They can be compared with information
extracted from helioseismology (see Christensen-Dalsgaard, these
proceedings). They can also be used to calculate the
expected fluxes of solar neutrinos which are
produced in the fusion of hydrogen into deuterium
($pp\rightarrow De^+\nu_e$ and  $pep\rightarrow D\nu_e$),
in electron capture by $^7$Be and in $\beta$ decay of $^8$B,
$^{13}$N, $^{15}$O and $^{17}$F using
the standard electroweak theory and the best available nuclear
reaction/decay rates.

Our calculations were performed with an updated version by Dar and Shaviv
(1994; 1995) of the solar evolution code of Kovetz and Shaviv (1994).
This updated version
includes  premain sequence evolution, diffusion of all elements,
partial ionization effects, and all the significant nuclear reactions
between the various elements which the sun is made of.
It uses updated values for the initial
solar element abundances, the solar age, the solar luminosity,
the nuclear reaction rates and the radiative opacities.
Neither nuclear equilibrium, nor complete ionization are assumed.
It employs a very fine zoning of the sun and accurate numerical
procedures to integrate the solar evolution equations from zero
age until the present day.
We refer the reader to its detailed description by Kovetz and Shaviv
(1994) and by Dar and Shaviv (1995). Here we highlight only
the choice of initial solar composition and nuclear reaction rates.

\subsection{  Initial Chemical Composition}
The initial element abundances influence significantly solar
evolution and the present density, chemical composition and temperature
in the solar core, which determine the solar neutrino fluxes.
In particular, the calculated radiative opacities, which determine the
temperature gradient in the solar interior, are very sensitive to the
abundances of the heavy elements (which are not fully
ionized in the sun).

Four major sources of information on the initial solar abundances were
used: The chemical composition of the most primitive class
of meteorites (type I carbonaceous chondrites), the solar photospheric
abundances, the chemical composition of the solar wind and the lunar
soil.

We have assumed that the initial solar
heavy metal abundances are given approximately by the meteoritic
(carbonaceous CI chondrites) values of Grevesse and Noels (1993a).
They better represent the initial values than
the photospheric ones, which have changed by mixing, convection and
diffusion during the solar evolution, and which are known
with much less accuracy.

The meteoritic CNO abundances are very much different from those
found in the solar photosphere. In particular, Carbon and Nitrogen are
underabundent in CI chondrites by about an order of magnitude
compared with their photospheric abundances (e.g., Sturenberg
and Howlweger 1990; Grevesse et al. 1990). Therefore, we have used
initial CNO abundances which yield their present
photospheric values when diffusion is included in the solar
evolution calculation.

The initial $^4$He mass fraction in the solar nebula which is known
only approximately, $0.24<Y<0.30$, has been treated
as an adjustable parameter.

The initial solar abundances of $^3$He and D, were taken to be
[$^3$He]/[H]=$(1.5\pm 0.3)\times 10^{-5}$ and
[D]/[H]=$(2.6\pm 1.0)\times 10^{-5}$, respectively. These values were
inferred by Geiss (1993) from analysis of solar wind particles
captured in foils exposed on the moon, from the lunar surface
 and from studies of primitive meteorites.

The photospheric abundances of $^7$Li, $^9$Be and $^{11}$B are smaller
by a factor of nearly 150, 3  and 10, respectively, than their
meteoritic abundances. The origin of such large differences is not
clear. However, the initial solar (meteoritic)
abundances of these elements are very
small and do not play any significant role in solar evolution
(their solar depletion, however, may store important information on the
true history of the sun).

\subsection{ Nuclear Reaction Rates}
The uncertainties in the nuclear reaction rates at solar conditions
are still large due to: (1) uncertainties in the measured cross sections
at laboratory energies, (2) uncertainties in their extrapolations
from laboratory energies down to solar energies,
(3) uncertainties in dense plasma effects (screening,
correlations and fluctuations) on reaction rates.
Neither the microscopic methods (for reviews see, e.g., Langanke
1991) such as the Resonating Group Method
(RGM) and  the Generator Coordinate Method (GCM), nor the potential
models such as the Optical Model (OM) and the Distorted Wave Born
Approximation (DWBA), can predict accurately enough the low energy
cross sections. Consider for instance the reaction $^7$Be(p,$\gamma)^8$B.
The RGM and the GCM, which are currently considered to be the
best theorerical methods for calculating direct nuclear reactions,
predict (see, e.g. Descouvemont and Baye 1994, Johnson et al. 1992)
 $S_{17}(0)\approx 25-30~eV\cdot b$.
However, a simple inspection of their predictions reveals that they
poorly reproduce the magnitude of the measured cross
section, the position of the resonance, the width of the resonance,
the height of the resonance and the observed shape of the cross section
as function of energy. To avoid these discrepancies
only the energy dependence of these models has been used
to extrapolate the measured cross  sections to $E=0$, yielding
(see, e.g., Johnson et al. 1992)  $S_{17}(0)\approx 22.4~eV\cdot b~.$
This value has been used in BP92 and BP95.
However, the procedure used by Johnson et al. (1992) is rather
ad hoc and questionable: (a) Their model does not reproduce
the measured energy dependence of the cross section at
lab energies. (b) The ``average cross section'' which they extrapolated
to solar energies was
obtained by averaging cross sections which differ by many standard
deviations and have different energy shapes (the cross sections measured
by Kavanagh (1960) and by Parker (1968) differ by more than $3\sigma$
from the cross sections
measured later by Vaughn (1970) and by Filippone (1983)
in the same energy ranges; ${\rm [Kavanagh]/
[Filippone]=1.34\pm 0.11}$, ${\rm [Parker]/[Vaughn]=1.42\pm 0.13}$;
see, e.g., Gai 1995).

Dar and Shaviv (1994) noted that sub-Coulomb radiative captures
and transfer reactions take place mainly when the colliding
nuclei are far apart. They argued that since
optical models describe well the shapes of the bound state
and relative motion wave functions outside the nuclear potential,
they should be preferred for extrapolating
the laboratory cross sections to solar energies. Alternatively,
they proposed to remove the trivial energy dependence due to
Coulomb barrier penetration to the effective distance R (calculated
or best fitted) where the reaction takes place and then to use
a simple polynomial fit to extrapolate the measured cross
sections to solar energies. The resulting astrophysical S factors,
which were obtained this way, are summarized in Table II. They include:

${\bf S_{17}(0)}$:
Extrapolations of the cross sections measured by Vaughn (1970) and
by Filippone(1983), using either simple potential models or the
very general properties of sub-Coulomb
cross section, gave $S_{17}(0)\approx 17~eV\cdot b$ (Barker and Spear
1986; Dar and Shaviv 1994; Kim et al. 1994).
Similar values were obtained also by different types of experiments:
Analysis of the virtual photodissociation reaction
$\gamma_v+^8$B$\rightarrow p+^7$Be
in the Coulomb field of Lead (Motobayashi et al. 1994) gave
$S_{17}(0)\approx 17~eV\cdot b$. A similar value,
$S_{17}(0)\approx 17.6~eV\cdot b$, was estimated from the virtual
reaction ${\rm p_v+^7Be\rightarrow B}$ measured
through the proton transfer reaction
$^3$He+$^7$Be$\rightarrow$D+$^8$B (Xu et al. 1994).
Consequently, we a have adopted the value $S_{17}=17~eV\cdot b$ in
our standard solar model calculations rather than the value
$S_{17}=22.4~eV\cdot b$ which was used in BP92 and BP95.

${\bf S_{34}(0)}$: The value
$S_{34}(0)= 0.51\pm 0.02~keV\cdot b $ was obtained from
measurements of the prompt $\gamma$-ray emitted in the reaction
${\rm ^3He+^4He}\rightarrow ^7Be+\gamma$, while measurements
of the induced $^7$Be activity led to a weighted average
$S_{34}(0)= 0.58\pm 0.02~keV\cdot b $ which is different by
3.5 standard deviations. The origin of this discrepancy is still
not known (Hilgemeier et al. 1988). Normalization to known cross
sections favour the lower value. Consequently, the value $S_{34}(0)=
0.45~keV\cdot b $ was  obtained by using the energy dependence of the
data of Krawinkel et al., 1982 to extrapolate the measured cross
sections for ${\rm ^3He+^4He\rightarrow ^7Be+\gamma}$
from prompt gamma ray emission
(Parker and Kavanagh 1963;
Osborne et al. 1982; Krawinkel et al. 1982 (multiplied by 1.4);
Hilgemeier et al. 1988) to solar energies. Note that the
$S_{34}(0)= 0.533~keV\cdot b $ was used in BP92 and
$S_{34}(0)= 0.524~keV\cdot b $ was used in BP95.

${\bf S_{33}(0)}$:
Extrapolation of the low energy data of Greife et al. (1994),
Krauss et al. (1987) and Dawarakanath and Winkler (1971)
on the reaction ${\rm ^3He+^3He\rightarrow ^4He+2p}$
gave $S_{33}(0)=5.6~MeV\cdot b$. Essentialy the same value,
$S_{33}(0)=5.57~MeV\cdot b$,
was obtained by Krauss et al. (1987)  and by Greife et al. (1944)
by applying a polynomial fit to their data.
The value $S_{33}(0)=5.0~MeV\cdot b$ was used in BP92 and
$S_{33}(0)=4.99~MeV\cdot b$ was used in BP95.

${\bf S_{11}(0)}$:
The reaction ${\rm p+p\rightarrow D+e^++\nu_e}$ has a cross section which is
too small to be measured directly in the laboratory.
 The weak isospin
related reactions
$\bar{\nu}_e+$D$\rightarrow e^+$+n+n, $\bar{\nu}_e+$D$\rightarrow\bar{
\nu}_e$+p+n, and $n+p\rightarrow D+\gamma $ have been measured
and can be used to obtain the relevant nuclear matrix element needed
for calculating the cross section for p+p$\rightarrow$D+$e^++\nu_e$.
This procedure yields a best value
$S_{11}(0)\approx 4.07\times 10^{-22}~keV\cdot b$
which is consistent with the value used by Caughlan and
Fowler (1988). It is 4.2\% larger than the value
$S_{11}(0)\approx 3.896\times 10^{-22}~keV\cdot b$ calculated recently
by Kamionkowski and Bahcall (1994).

{\bf Screening Enhancement of Reaction Rates}:
Screening of target nuclei by  electrons known to enhance
significantly laboratory nuclear cross sections at very low
energies (e.g., Engstler et al. 1988).
A complete theoretical understanding of the effect
is still lacking (see, e.g.,
Shoppa et al. 1993; Rolfs 1994 and references therein).
The screening enhancement factors of the nuclear reaction rates
near the center of the sun (Bahcall 1989) are quite
considerable, being 5\%, for the pp and pep reactions, 20\% for
the ${\rm ^3He^3He}$, ${\rm ^3He^4He}$ and ${\rm p^7Be}$ reactions,
and 30\%, 35\% and 40\% for the p capture by the C, N and O isotopes,
respectively. However, these screening enhancement factors are based
on the Debye-Huckel approximation which is not valid for the
conditions prevailing near the center of the sun.

Reliable estimates of the nuclear reaction rates in dense plasma require
numerical N body simulations of the behaviour of electrons and ions in
a dense plasma. Such simulations can be used to evaluate the effects
of screening, correlations and fluctuations on the thermonuclear
reaction rates in dense plasmas (Shaviv and Shaviv 1995). They have
not yet been incorporated in our solar evolution code.

However, to
test the sensitivity of the standard solar models to the screening
corrections we have carried out the calculations with, and without,
the standard screening enhancement factors of all the thermonuclear
reaction rates. We have found (Dar and Shaviv 1994) that
removing/including the screening enhancement factors for
{\bf all} nuclear reaction rates
had only a small net effect on the calculated solar neutrino
fluxes, due to accidental cancellations.
Screening factors may, however, play important role if their ratios
for the different reactions are changed.

\subsection{ Radiative Opacities}
The radiative opacities depend on the local chemical composition,
density and temperature in the sun. We have used radiative opacity tables
computed by the OPAL group at Lawrence Livermore National Laboratory
that were kindly provided to us by
F.J. Rogers. They are updated version of the OPAL tables of
Rogers and Iglesias (1992) for the most recent determination
by Grevesse and Noels (1993a)
of the heavy element composition of the sun from
the meteoritic and photospheric data.

\subsection{ Diffusion}
The Kovetz-Shaviv code calculates the diffusion of all the individual
elements from the premain sequence phase to the present age. The binary
and thermal diffusion coefficients depend on the squared ionic charges.
The ionization state of each element in every shell is
calculted by solving the Saha equations for all the
elements in each shell. All elements with mass fractions less than
$10^{-5}$ are treated as trace elements. Their collisions with other
trace elements are neglected.

As a consequence of diffusion the surface and internal element
abundances change for each element in a slightly different way. Diffusion
depletes the surface abundances of $^4$He and the heavy elements.
For C, N and O only their present photospheric abundances are known.
Consequently, in our solar model calculations
their initial solar abundances were adjusted to
reproduce their observed photospheric abundances.
For all other elements the initial meteoritic abundances were used
to predict their final surface abundances which can be compared
with their observed photospheric abundances.
The unknown initial abundance of $^4$He was treated as a free parameter.
It was adjusted to best reproduce the presently observed sun.
Its predicted surface abundance today can be compared with the value
derived from helioseismology (Hernandez and Christensen - Dalsgaard
1994).

\section{ Results and Comparison With Observations}
Our predictions of the solar neutrino fluxes for three standard
solar models are summarized in Table IIIa  and are compared
with the results from the four solar neutrino experiments.
The three models differ only
in their teatment of element diffusion: The model
labeled DSND does not include element diffusion. The models
labeled DS94 and DS95 include element diffusion. Model
DS94 assumes that the initial heavy metal abundances in the sun
were equal to their meteoritic values, and those of C, N and O
were equal to their observed photospheric abundances, as summarized
in Table I. In model DS95 the initial abundances of C, N and O were
adjusted to yield their
present day photospheric abundances while the heavy metal abundances
were assumed to be those found in primitive meteorites, as
summarized in Table I. Table IIIb presents some physical
characteristics of the three models.
As can be seen from Table IIIa, all three
models yield a $^8$B solar neutrino flux consistent with that measured
by Kamiokande. However all three models predict capture rates
in the Chlorine and Gallium experiments which are significantly larger
than those measured by Homestake, GALLEX and SAGE
(we have used the neutrino  cross sections from Table 8.2  of
Bahcall (1989) to convert solar neutrino fluxes to capture rates).

Table IV present a comparison between four solar models. The model
labeled BP92 is the best model of BP92. It includes diffusion of
protons and $^4$He but not of other elements. The model
labeled BP95 is the best model of BP95 which includes also diffusion
of the heavy metals but assumes that all the heavy elements diffuse
like fully ionized iron. The predictions of the DS models differ
significantly from the BP models because of different
input physics, approximations and numerical methods. Most of the
differences are due to the use of different reaction rates
as summarized in Table II and the different treatments of diffusion.
This is demonstrated in Table V where we present a comparison
between the best model of BP95 without diffusion labeled BP95ND,
and a solar model, labeled DS(BPND) calculated with the Kovetz-Shaviv
stellar evolution code with the same physical and astrophysical input
parameters and the same nuclear reaction rates used in BP95ND,
and without inclusion
of element diffusion. As can be seen from Table V the two
calculations yield similar results. Even the fluxes  of $^8$B and CNO
solar neutrinos which, under the imposed solar boundary conditions,
are very sensitive to the central solar temperature
differ by less than 4\%.
These remaining  differences are probably due to the use of
different equations of state, numerical methods,
fine zoning and time steps in the two codes and
due to the inclusion of premain sequence evolution in our code.

To emphasize the important role that might be played by diffusion,
Table V  also include  the current best solar models of Dar and
Shaviv (1995) and of Bahcall and Pinsonneault (1995), which include
diffusion of all elements. As can be seen from Tables V, Bahcall
and Pinsonneault (1995) found  rather large increases in their predicted
$^7$Be, $^8$B, $^{13}$N, $^{15}$O and $^{17}$F
solar neutrino fluxes; 14\%, 36\%, 52\%, 58\%, and 61\%,
respectively, compared with their model (BP95ND) with no diffusion.
These induce 36\%, 33\%, 9\% increases in the predicted
rates in Kamiokande, Homestake, and in GALLEX and SAGE, respectively.
However, we predict more moderate increases due to diffusion,
4\%, 10\%, 23\%, 24\% and 25\%, respectively, in the above fluxes,
which correspond to 10\%, 10\% and 2\% increases in the predicted
rates in Kamiokande, Homestake, and in GALLEX and SAGE, respectively.
The differences in the effects of diffusion in DS94 and BP95
are mainly due to two
reasons: (a) In the calculations of Bahcall and Pinsonneault (BP95)
all heavy elements were assumed to diffuse at the same rate as
fully ionized iron while in the Dar-Shaviv calculations (DS94)
followed the diffusion of all the elements separately and
used diffusion coefficients for the actual ionization state of each
element.
(b) Bahcall and Pinsonneault assumed that the meteoritic
abundances represent the solar surface abundances today and not
their initial values. They adjusted their initial values to
reproduce  surface abundances today equal to the meteoritic values.
Thus they have actually used an initial ratio
$Z/X=0.0285$ (see Table IV of BP95)
while the observed ratio in meteorites is $Z/X= 0.0245$ (Grevesse
and Noels 1993). Dar and Shaviv (1994) used the meteoritic values
for the initial metalic abundances and predicted present day
depleted surface abundances. Unfortunately, the uncertainties in the
measured photospheric CNO abundances (typically 30\%-40\%) are much
larger than their predicted depletion (typically 10\%) and do not
allow a reliable test of the predicted photospheric abundances.

Standard solar models do not explain the large depletion in the abundance
of Li in the solar photosphere compared to its meteoritic abundance.
Such a depletion may indicate significant mixing of surface
material to solar depths where Li is burned efficiently.
Mixing may , however, inhibit the inward diffuson of
the heavy elements. Thus, it is not clear whether solar models with
diffusion provide a more realistic description of the
sun than solar models without diffusion, although
helioseismology data is better explained by SSM's with
diffusion. See, e.g., Cristensen-Dalsgaard et al. 1993.

\section{Where Are The $^7$Be Neutrinos ? }

Electron capture by $^7$Be into the ground state of $^7$Li
produces 862 keV neutrinos. The threshold
energy for neutrino absorption by $^{37}$Cl is 814 keV. Thus, absorption
of $^7$Be neutrinos by $^{37}$Cl produces 48 KeV electrons.
The average energy of the pp solar neutrinos is 265 KeV. The
threshold energy for neutrino absorption in $^{71}$Ga is 233 KeV.
Consequently, the produced electron has a typical energy of
33 keV. The de Broglie wave lengths of such electrons are
larger than the Bohr radii of the atomic K shells in Cl and Ga
and their energies are similar to the kinetic energies of electrons
in the K shells. Consequently, screening of the nuclear
charge by atomic electrons and final state interactions (exchange
effects, radiative corrections, nuclear recoil against the electronic
cloud, etc.) may slightly reduce the absorpton cross sections of
pp neutrinos in $^{71}$Ga (perhaps making room for the expected
contribution of $^7$Be in Gallium ?)
and of $^7$Be neutrinos in $^{37}$Cl (perhaps making the solar neutrino
observations of Kamiokande and the Homestake experiment compatible).
It is interesting
to note that although final state interactions in Tritium beta
decay have been studied extensively, they do not explain yet why its
end-point spectrum ($E_e\sim 18.6~kEV$)
yields consistently, in all recent measurements,
a negative value for the squared mass of the electron neutrino.
Final state interactions in $^{37}$Cl and $^{71}$Ga are expected
to be much larger because of their much larger values of Z.
Note also that the above explanation implies that experiments like
BOREXINO and HELLAZ will observe the full $^7$Be solar neutrino flux
while the MSW solution predicts that it will be strongly suppressed.

Even if the $^7$Be solar neutrino flux is strongly suppressed,
it does not elliminate yet standard physics solutions to the
solar neutrino problem. For instance,
collective plasma physics effects, such as
very strong magnetic or electric fields near the center of the sun,
may polarize  the plasma electrons, and affect the
branching ratios of electron capture by $^7$Be (spin $3/2^-$) into
the ground state
(spin $3/2^-$,  $E_{\nu_e}=0.863~MeV$, BR=90\% and the excited state
(spin $1/2^-$,  $E_{\nu_e}=0.381~MeV$, BR=10\%) of $^7$Li. Since
solar neutrinos with $E_{\nu_e}=0.381~MeV$ are below the threshold (0.81
MeV) for capture in $^{37}$Cl and have a capture cross section in
$^{71}$Ga that is smaller by about a factor of 6 relative to
solar neutrinos with $E_{\nu_e}=0.863~MeV$, therefore a large
suppression in the branching ratio to the ground state can produce
large suppressions of the $^7$Be solar neutrino signals in $^{37}$Cl
and in $^{71}$Ga.

\section{\bf Conclusions:}
Solutions to the solar neutrino problem which
do not invoke physics beyond the standard electroweak model
are not ruled out yet:

The solar neutrino problem may be a terrestrial
problem. The neutrino capture cross sections near threshold
in the radiochemical experiments may be different
from the calculated cross sections.
The inferred solar neutrino fluxes
from the GALLEX and HOMESTAKE experiments may be different from
the true solar neutrino fluxes. They do
not established beyond doubt that the $^7$Be solar neutrino
flux is strongly suppressed.
BOREXINO and HELLAZ may be able to do that.

The solar neutrino problem may be a problem of the standard
solar model. Namely, the model may not provide yet an accurate enough
description of the sun and the nuclear reactions that take place in it.
The deviations of the experimental results from those predicted by the
standard solar models may reflect the approximate
nature of the standard solar models (which neglect, or treat
only approximately, many effects and do
not explain yet solar activity nor the surface depletion of
Lithium, Berilium and Boron relative to their meteoritic values,
that may or may not be relevant to the solar neutrino problem).
Improvements of the standard solar model should continue.
In particular,
dense plasma effects on nuclear reaction rates and radiative opacities,
which are not well understood, may strongly affect the SSM predictions
and should be further studied, both theoretically and experimentally.
Relevant information may be obtained from studies of thermonuclear
plasmas in inertial confinement experiments. Useful information
may also be obtained from improved data on screening effects
in low energy nuclear cross sections of ions, atomic beams and molecular
beams incident on a variety of gas, solid and plasma targets.

Better knowledge of low energy nuclear cross sections is
badly needed. Measurement of crucial low energy nuclear cross sections
by new methods, such as measurements of the
cross sections for the radiative captures ${\rm p+^7Be\rightarrow
^8B+\gamma}$ and ${\rm ^3He+^4He\rightarrow ^7Be+\gamma}$ by
photodissociation of $^8$B  and $^7$Be in the coulomb field of
heavy nuclei are highly desireable.

The $^{37}$Ar production rate in $^{37}$Cl may indeed be smaller
than that expected from the total solar neutrino flux measured
by electron scattering in the Kamiokande experiment. In that case
neutrino oscillations, and in particular the MSW effect, may
be the correct solution to the solar neutrino problem. Only future
experiments like SNO and Superkamiokande will be able to supply
a definite proof that Nature has made use of this beautiful effect.
\section{Acknowledgement}
This research was supported in part by the
Technion Fund For Promotion of Research.

\clearpage

\def\reference{\item}

\begin{enumerate}

\reference\noindent

\reference\noindent
Anders, E. and Grevesse, N., 1989, Geochim. Cosmochim. Acta,
{\bf 53}, 197

 \reference\noindent
Anselmann, P. et al., 1995,  Phys. Lett. B {\bf 342}, 440

 \reference\noindent
Bahcall, J.N. 1989, "Neutrino Astrophysics", (Cambridge
University Press 1989).

 \reference\noindent
Bahcall, J.N., 1994, Phys. Lett. B {\bf 338}, 276

 \reference\noindent
Bahcall, J.N., 1995, Nucl. Phys. B (Proc. Suppl.) {\bf 38}, 98

 \reference\noindent
Bahcall, J.N. and Bethe, H. 1991, Phys. Rev. D. {\bf 44}, 2962

 \reference\noindent
Bahcall, J.N. and Pinsonneault, M. 1992, Rev. Mod. Phys. {\bf 64}, 885

 \reference\noindent
Bahcall, J.N. and Pinsonneault, M. 1995, Rev. Mod. Phys., (submited)

 \reference\noindent
Bahcall, J.N. and Ulrich, R.K., 1988, Rev. Mod. Phys. {\bf 60}, 297

 \reference\noindent
Barker, F.C. and Spear, R.H. 1986, ApJ. {\bf 307}, 1986

 \reference\noindent
Berezinsky, V., 1994, Comm. Nucl. Part. Phys. {\bf 21}, 249

 \reference\noindent
Castellani, V. et al., 1994, Phys. Lett. B {\bf 324}, 425

 \reference\noindent
Caughlan, G.R. and Fowler, W.A. 1988, Atomic and Nucl. Data
Tables {\bf 40}, 283

 \reference\noindent
Christensen-Dalsgaard, J., 1994, Europhys. News {\bf 25}, 71

 \reference\noindent
Christensen-Dalsgaard, J. and
Dappen, W. 1992, A\&A Rev. {\bf 4}, 267

 \reference\noindent
Christensen-Dalsgaard, J. et al., 1993, ApJ. {\bf 403}, L75

 \reference\noindent
Clayton, D. 1968, Principles of Stellar Evolution \& Nucleosyn.
(McGraw-Hill)

 \reference\noindent
Cleveland, B.T. et al., 1995, Nucl. Phys. B (Proc. Suppl.) {\bf 38}, 47

 \reference\noindent
Dar, A. and Shaviv G. 1994, Proc. VI Int. Conf. on Neutrino Telescopes
(edt. M. Baldo-Ceolin) p. 303. See also Shaviv G., 1994

 \reference\noindent
Dar, A., and Shaviv, G., 1995, ApJ, in press.

 \reference\noindent
Dawarakanath, M.R., and Winkler H.,  1971, Phys. Rev. C {\bf 4}, 1532

 \reference\noindent
Descouvemont and Baye, 1994, Nucl. Phys. {\bf A567}, 341

 \reference\noindent
Dzitko, H. et al., 1995, ApJ. {\bf 447}, 428

 \reference\noindent
Engstler, S. et al, 1988, Phys. Lett. B {\bf 202}, 179

 \reference\noindent
Filippone, B.W. et al. 1983, Phys. Rev. C {\bf 28}, 2222

 \reference\noindent
Gai, M., 1995,  Nucl. Phys. B (Proc. Suppl.) {\bf 38}, 77

 \reference\noindent
Geiss, J.,  1993  in Origin and Evolution of the Elements, ed.
N. Prantzos et al (Cambridge Univ. Press, Cambridge) p. 89

 \reference\noindent
Greife, U. et al., 1994, Nucl. Inst. \& Methods. A {\bf 350}, 327

 \reference\noindent
Grevesse, N., 1991, A\&A, {\bf 242}, 488

 \reference\noindent
Grevesse, N. and Noels, A., 1993, in {\it Origin and Evolution of the
Elements, edsN. Prantzos et al.} (Cambridge Univ. Press) p. 15

 \reference\noindent
Grevesse, N. and Noels, A., 1993, Phys. Scripta {\bf T47} , 133

 \reference\noindent
Hata, N. et al., 1994, Phys. Rev. D {\bf 49}, 3622

 \reference\noindent
Hata, N. and Langacker, P., 1995, Phys. Rev. D {\bf 52}, 420

 \reference\noindent
Hernandez, E.P. and Christensen-Dalsgaard, J., 1994, MNRAS
{\bf 269}, 475

 \reference\noindent
Hilgemeier, M. et al., 1988, Z. Phys. A {\bf 329}, 243

 \reference\noindent
Johnson, C.W. et al. 1992, ApJ. {\bf 392}, 320

 \reference\noindent
Kamionkowski, M. and Bahcall, J., (1994), ApJ. {\bf 420}, 884

 \reference\noindent
Kavanagh, R.W., et al., 1969, Bull. Am. Phys. Soc. {\bf 14}, 1209

 \reference\noindent
Kim, Y.E., et al., 1995,  Nucl. Phys. B (Proc. Suppl.) {\bf 38}, 293

 \reference\noindent
Kovetz, A. and Skaviv, G., 1994, ApJ. {\bf 426}, 787

Krauss, A. et al., 1987, Nucl. Phys. A {\bf 467}, 273

 \reference\noindent
Krawinkel, H. et al., 1982, Z. Phys. A, {\bf 304}, 307

 \reference\noindent
Kwong, W. and Rosen, S, 1994, Phys. Rev. Lett. {\bf 73}, 369

 \reference\noindent
Langanke, K. 1991, in Nuclei in The Cosmos (ed. H, Oberhummer) Berlin:
Springer-Verlag, p. 61

 \reference\noindent
Mikheyev, P. and Smirnov, A. Yu. 1985, Yad. Fiz. {\bf 42}, 1441

 \reference\noindent
Motobayashi, T., et al., 1994, Phys. Rev. Lett., {\bf 73}, 2680

 \reference\noindent
Osborne, J.L. et al., 1984, Nucl. Phys. {\bf A419}, 115

 \reference\noindent
Parke, S.J. 1995, Phys. Rev. Lett. {\bf 74}, 839

 \reference\noindent
Parker, P.D., 1966, Phys. Rev. Lett. {\bf 150}, 851

 \reference\noindent
Parker, P.D., 1968, ApJ. {\bf 153}, L85

 \reference\noindent
Parker, P.D., and Kavanagh, R. W., 1963 Phys. Rev. {\bf 131}, 2578

 \reference\noindent
Rogers, F.J. and Iglesias, C.A., 1992, ApJ. Suppl. {\bf 79}, 57

 \reference\noindent
Rolfs, C., 1994  Nucl. Phys. B (Proc. Suppl.) {\bf 35} 334

 \reference\noindent
Sackman, I.J. et al., 1990, ApJ. {\bf 360}, 727

 \reference\noindent
Schramm, D.N. and Shi, X, 1994, Nucl. Phys. B (Proc. Suppl.){\bf 35}, 321

 \reference\noindent
Shaviv, G. 1995, Nucl. Phys. B (Proc. Suppl.) {\bf 38}, 81

 \reference\noindent
Shaviv, G. and Shaviv, N., 1995, submitted to ApJ.

 \reference\noindent
Shi, X.D. et al., 1994, Phys. Rev. D {\bf 50}, 2414

 \reference\noindent
Shoppa, T.D. et al., 1993, Phys. Rev. C {\bf 48}, 837

 \reference\noindent
Sturenburg, S. and Holweger, H., 1990, A\&A {\bf 237}, 125

 \reference\noindent
Turck-Chieze, S. et al., 1988, ApJ. {\bf 335}, 415

 \reference\noindent
Turck-Chieze, S. and Lopes, I., 1993, ApJ. {\bf 408}, 347

 \reference\noindent
Vaughn et al., 1970, Phys. Rev. {\bf C2}, 1657

 \reference\noindent
Volk, H. et al., 1983, Z. Phys. A {\bf 310}, 91

 \reference\noindent
Wolfenstein, L. 1978, Phys. Rev. {\bf D17}, 2369

 \reference\noindent
Wolfenstein, L. 1979, Phys. Rev. {bf D20}, 2634

\end{enumerate}
 \clearpage
{
{\bf Table I:}  Summary of Information
on Abundances of Various Elements relative to Hydrogen
( A$\equiv$log([A]/[H])+12) in  Primitive Meteorites, in the Solar
Photosphere, in the Solar Wind and in the Local Interstellar Medium,
Used In The DS Standard Solar Models.}

{$$\matrix{{\rm Element}&{\rm Abundance}&{\rm Source}&{\rm Reference}\cr
{\rm D }&7.22\pm 0.05&{\rm LISM,~Meteorites} & {\rm Linsky~1993,~
Geiss~1993}\cr
^{\rm 3}{\rm He}&7.18\pm0.08&{\rm Meteorites,~Solar~Wind}&{\rm
Geiss~1993}\cr
^{\rm 7}{\rm Li}&1.54\pm
0.0X&{\rm Meteorites} & {\rm Anders~and~Grevesse~1989}\cr
^{\rm 9}{\rm Be}&1.13\pm
0.0X&{\rm Meteorites} & {\rm  Anders~and~Grevesse~1989}\cr
^{\rm 12}{\rm C}&8.55\pm
0.05&{\rm Photosphere} & {\rm Grevesse~and~Noels~1993}\cr
^{\rm 13}{\rm C}&6.60\pm
0.05& {\rm Photosphere} & {\rm Grevesse~and~Noels~1993}\cr
^{\rm 14}{\rm N}&7.97\pm 0.07&
{\rm Photosphere }& {\rm Grevesse~and~Noels~1993}\cr
^{\rm 16}{\rm O}&8.78\pm 0.07&
{\rm Photosphere } & {\rm Grevesse~and~Noels~1993}\cr
^{\rm 20}{\rm Ne}&8.08\pm 0.06&
{\rm Photosphere } & {\rm Grevesse~and~Noels~1993}\cr
^{\rm 23}{\rm Na}&6.33\pm 0.03&
{\rm Meteorites~and~Photosphere } & {\rm Grevesse~and~Noels~1993}\cr
^{\rm 24}{\rm Mg}&7.58\pm 0.05&" &"\cr
^{\rm 27}{\rm Al }&6.47\pm 0.07&"&"\cr
^{\rm 28}{\rm Si}&7.66\pm 0.05&"&"\cr
^{\rm 31}{\rm P}&5.45\pm 0.04&"&"\cr
^{\rm 32}{\rm S}&7.21\pm 0.06&"&"\cr
^{\rm 35}{\rm Cl }&5.5\pm 0.3&"&"\cr
^{\rm 40}{\rm Ar}&6.52\pm 0.1&"&"\cr
^{\rm 40}{\rm Ca}&6.36\pm 0.02&"&"\cr
^{\rm 40}{\rm K}&4.85&"&"\cr
^{\rm 45}{\rm Sc}&3.08&"&"\cr
^{\rm 48}{\rm Ti}&5.02\pm 0.06&"&"\cr
^{\rm 50}{\rm V}&3.99&"&"\cr
^{\rm 52}{\rm Cr}&5.67\pm 0.03&"&"\cr
^{\rm 55}{\rm Mn}&5.39\pm 0.03&"&"\cr
^{\rm 56}{\rm Fe}&7.50\pm 0.04&"&"\cr
^{\rm 63}{\rm Cu}&4.15&"&"\cr
^{\rm 58}{\rm Ni}&6.25\pm 0.04&"&"\cr
^{\rm 64}{\rm Zn}&4.33&"&"\cr
\rm Z\rm /\rm X&\rm 0.0245&"&"\cr}$$}

\clearpage
{\bf Table II:}
Comparison Between the Astrophysical S Factors
for the pp-chain Reactions used in BP95 and in DS94 and DS95.
The values of S are given in $keV\cdot b$ Units.
 \bigskip

$$\matrix{ {\rm Reaction } \hfill & {\rm {S}^{BP}}\rm (0) \hfill &{\rm
S}^{DS}\rm (0)
 \hfill \cr
^{\rm 1}{\rm H}(\rm p\rm ,{\rm e}^{\rm +}{\rm \nu }_{e}{\rm ) D}
\hfill &
\rm 3.896\times {10}^{-22} \hfill &\rm 4.07\times {10}^{-22} \hfill \cr
^{\rm 1}{\rm H}({\rm p}{e}^{\rm -}{\rm \nu}_{e} ){\rm D} \hfill &
{\rm Bahcall~89 } \hfill &{\rm CF88} \hfill \cr
^{\rm 3}{\rm He}(^{\rm 3}{\rm He},2\rm p\rm ){\rm He}^{\rm 4} \hfill &
4.99\times {10}^{3} \hfill &5.6\times {10}^{3} \hfill \cr
^{\rm 3}{\rm He}(^{\rm 4}{\rm He},\rm \gamma \rm )^{\rm 7}{\rm Be}
\hfill &0.524 \hfill & 0.45 \hfill \cr
^{\rm 7}{\rm Be}({\rm e}^{\rm -},{\rm \nu }_{e}{\rm )}{\rm Li}^{\rm 7}
\hfill & {\rm Bahcall~89} \hfill &{\rm CF88} \hfill \cr
^{\rm 7}{\rm Be}(\rm p\rm ,\rm \gamma \rm )^{\rm 8}{\rm B} \hfill &0.0224 \hfill
& 0.017 \hfill \cr}$$

\bigskip
{\bf Table IIIa:} Comparison Between Solar Neutrino Fluxes
Predicted by the current best Standard Solar Models of Dar and
Shaviv, with and without
element diffusion, and the Solar Neutrino Observations.
\bigskip
$$\matrix{\nu~{\rm Flux}\hfill & {\rm  DSND} \hfill & {\rm  DS94} \hfill
& {\rm   DS95}
\hfill &{\rm  Observation} \hfill &{\rm
Experiment} \hfill \cr
{\rm \phi }_{\nu }(pp)~[{10}^{10}{cm}^{-2}{s}^{-1}] \hfill &6.10
\hfill &6.06 \hfill &6.03 \hfill &\mit &\rm \cr
{\phi }_{\nu }(pep)~[{10}^{8}{cm}^{-2}{s}^{-1}] \hfill &
1.43\hfill &1.42\hfill&1.40\hfill &\rm \hfill &\rm \cr
{\phi }_{\nu }(^{7}{Be})~[{10}^{9}{cm}^{-2}{s}^{-1}]\hfill &4.03\hfill
&4.00\hfill &4.20\hfill &\rm \ll \phi_{\nu}^{SSM}(^7Be)\hfill
&\rm ALL\cr
{\phi }_{\nu }(^{8}{B})~[{10}^{6}{cm}^{-2}{s}^{-1}]\hfill &2.54\hfill
&2.60\hfill &2.87\hfill &\rm 2.9\pm 0.4\hfill &\rm Kamiokande\cr
{\phi }_{\nu }(^{13}{N})~[{10}^{8}{cm}^{-2}{s}^{-1}]\hfill &3.21\hfill
&3.30\hfill
&3.94\hfill &\rm \hfill &\rm \cr
{\phi }_{\nu }(^{15}{O})~[{10}^{8}{cm}^{-2}{s}^{-1}]\hfill &3.13\hfill
&3.19\hfill
&3.88\hfill &\rm \hfill &\rm \cr
{\phi }_{\nu }(^{17}{F})~[{10}^{6}{cm}^{-2}{s}^{-1}]\hfill &3.77\hfill
&3.84\hfill
&4.71\hfill &\rm \hfill &\rm \cr
\Sigma (\phi \sigma)_{Cl}~[SNU]\hfill &\rm 4.2\pm 1.2\hfill &
\rm 4.2\pm 1.2\hfill &\rm 4.6\pm 1.6\hfill&\rm 2.55\pm 0.25
 \hfill &{\rm Homestake} \cr
\Sigma(\phi\sigma)_{Ga}~[SNU]\hfill &\rm 116\pm 6\hfill
&\rm 116\pm 6 \hfill &\rm 119\pm 7\hfill
&\rm 79\pm 12 \hfill &{\rm GALLEX}\cr
\Sigma(\phi\sigma)_{Ga}~[SNU]\hfill &\rm 116\pm 6\hfill
&\rm 116\pm 6 \hfill &\rm 119\pm 7\hfill
&\rm 74\pm 16 \hfill &{\rm SAGE}\cr}$$

\vfill
\eject

{\bf Table IIIb:} Characteristics of the DS
Solar Models in Table IIIa (c=center; s=surface;
bc=base of convective zone; $\bar N=log([N]/[H])+12$).

$$\matrix{{\rm Parameter}\hfill& {\rm DSND}\hfill& {\rm DS94}\hfill& {\rm
DS95}\hfill \cr
{T}_{c}~[{10}^{7}K] \hfill &1.553 \hfill &1.554 \hfill &1.562 \hfill \cr
{\rho }_{c}~[g~c{m}^{-3}]\hfill&154.9 \hfill&155.3\hfill&155.9\hfill
& &\cr
{X}_{c}\hfill&0.3491\hfill &0.3462\hfill &0.3381 \hfill \cr
{Y}_{c}\hfill&0.6333 \hfill &0.6359 \hfill &0.6425 \hfill \cr
{Z}_{c}\hfill&0.01757 \hfill &0.01802 \hfill
&0.01940 \hfill \cr
{X}_{s}\hfill&0.6978 \hfill &0.7243 \hfill &0.7171 \hfill \cr
{Y}_{s}\hfill&0.2850 \hfill &0.2597 \hfill &0.2658 \hfill \cr
{Z}_{s}\hfill&0.01703 \hfill &0.01574 \hfill
&0.01696 \hfill \cr
\overline{N}_s{(^{12}C})\hfill&8.55\hfill &8.50 \hfill&8.55 \hfill \cr
\overline{N}_s{(^{14}N})\hfill&7.97\hfill &7.92 \hfill&7.97 \hfill \cr
\overline{N}_s{(^{16}O})\hfill&8.87\hfill &8.82 \hfill&8.87 \hfill \cr
\overline{N}_s{(^{20}Ne})\hfill&8.08\hfill &8.03 \hfill&8.08 \hfill \cr
\overline{X}_s{(\geq^{24}Mg})\hfill&0.00464
\hfill &0.00414 \hfill &0.00415 \hfill \cr
{R}_{conv}~[R/R_{\odot}]\hfill&0.7306 \hfill &0.7105 \hfill &0.7066
\hfill \cr
{T}_{bc}~[{10}^{6}{\rm K}]\hfill&1.97 \hfill &2.10 \hfill &2.12
\hfill \cr
{T}_{eff}~[{\rm K}]\hfill&5895 \hfill &5920 \hfill &5919
\hfill \cr}$$

\bigskip
{\bf Table IVa:} Comparison Between Solar Neutrino Fluxes
Predicted by the Dar-Shaviv  Models and the
Bahcall-Pinsonneault best Solar Models.
\bigskip
$$\matrix{\rm & {\rm BP92}& {\rm  DS94} & {\rm BP95}& {\rm
DS95}\cr
{\phi }_{\nu }(pp)~[{10}^{10}{cm}^{-2}{s}^{-1}]\hfill &
6.00\hfill &6.06\hfill &5.91\hfill &6.03\cr
{\phi }_{\nu }(pep)~[{10}^{8}{cm}^{-2}{s}^{-1}]\hfill &
1.43\hfill & 1.42\hfill &1.40\hfill &1.40\cr
{\phi }_{\nu }(^{7}{Be})~[{10}^{9}{cm}^{-2}{s}^{-1}]\hfill &
4.89\hfill &4.00\hfill &5.15\hfill &4.20\cr
{\phi }_{\nu }(^{8}{B})~[{10}^{9}{cm}^{-2}{s}^{-1}]\hfill &
5.69\hfill &2.60\hfill &6.62\hfill &2.87\cr
{\phi }_{\nu }(^{13}{N})~[{10}^{9}{cm}^{-2}{s}^{-1}]\hfill &
4.92\hfill &3.30\hfill &6.18\hfill &3.94\cr
{\phi }_{\nu }(^{15}{O})~[{10}^{9}{cm}^{-2}{s}^{-1}]\hfill &
4.26\hfill &3.19\hfill &5.45\hfill &3.88\cr
{\phi }_{\nu }(^{17}{F})~[{10}^{9}{cm}^{-2}{s}^{-1}]\hfill &
5.39\hfill &3.84\hfill &6.48\hfill &4.71\cr
\Sigma (\phi \sigma)_{Cl}~[SNU]\hfill &\rm 8\pm 1\hfill &
\rm 4.2\pm 1.2\hfill
&\rm 9.3\pm 1.4\hfill&\rm 4.6\pm 1.6\hfill \cr
\Sigma(\phi\sigma)_{Ga}~[SNU]\hfill &\rm 132\pm 7\hfill
&\rm 116\pm 6\hfill &\rm 137\pm 8\hfill
&\rm 119\pm 7\hfill \cr}$$

\vfill
\eject
{\bf Table IVb} Characteristics of the BP95, DS94, and DS95
Solar Models in Table IIIa
(c=center; s=surface; bc=base of convective zone;
${\rm \bar N=log([N]/[H])+12)}$.
\bigskip
$$\matrix{{\rm Parameter}\hfill& {\rm BP95}\hfill& {\rm DS94}\hfill& {\rm
DS95}\hfill \cr
{T}_{c}~[{10}^{7}K] \hfill &1.584 \hfill &1.554 \hfill &1.562 \hfill \cr
{\rho }_{c}~[g~c{m}^{-3}]\hfill&156.2 \hfill&155.3\hfill&155.9\hfill \cr
{X}_{c}\hfill&0.3333\hfill &0.3462\hfill &0.3381 \hfill \cr
{Y}_{c}\hfill&0.6456 \hfill &0.6359 \hfill &0.6425 \hfill \cr
{Z}_{c}\hfill&0.0211\hfill&0.01802 \hfill
&0.01940 \hfill \cr
{X}_{s}\hfill&0.7351 \hfill &0.7243 \hfill &0.7171 \hfill \cr
{Y}_{s}\hfill&0.2470 \hfill &0.2597 \hfill &0.2658 \hfill \cr
{Z}_{s}\hfill&0.01798 \hfill &0.01574 \hfill
&0.01696 \hfill \cr
\overline{N}_s{(^{12}C})\hfill&8.55\hfill &8.50 \hfill&8.55 \hfill \cr
\overline{N}_s{(^{14}N})\hfill&7.97 \hfill &7.92 \hfill&7.97\hfill \cr
\overline{N}_s{(^{16}O})\hfill&8.87 \hfill &8.82 \hfill&8.87 \hfill \cr
\overline{N}_s{(^{20}Ne})\hfill&8.08 \hfill &8.03 \hfill&8.08 \hfill \cr
\overline{X}_s{(\geq^{24}Mg})\hfill&  &0.00414 \hfill&
0.00415 \hfill \cr
{R}_{conv}~[R/R_{\odot}]\hfill&0.712 \hfill &0.7105 \hfill &0.7066
\hfill \cr
{T}_{bc}~[{10}^{6}{\rm K}]\hfill&2.20 \hfill &2.10 \hfill &2.12
\hfill \cr
{T}_{eff}~[{\rm K}]\hfill& \hfill &5920 \hfill &5919
\hfill \cr}$$

\bigskip
{\bf Table V:} Comparison between the solar neutrino fluxes
calculated from the best standard solar model with no diffusion of BP95
and those calculated with the Dar-Shaviv SSM code with the same
nuclear reaction rates, opacities, composition and astrophysical
parameters. The predictions of the current best standard solar models of
Dar and Shaviv and of Bahcall and Pinsonneault are also included.
\bigskip
$$\matrix{\rm  & {\rm BP95(ND)}& {\rm  DS(BPND)} & {\rm BP95}& {\rm
DS95}\cr
{\phi }_{\nu }(pp)~[{10}^{10}{cm}^{-2}{s}^{-1}]\hfill
&6.01\hfill &6.08\hfill &5.91\hfill &6.03\cr
{\phi }_{\nu }(pep)~[{10}^{8}{cm}^{-2}{s}^{-1}]\hfill
&1.44\hfill &1.43\hfill &1.40\hfill &1.40\cr
{\phi }_{\nu }(^{7}{Be})~[{10}^{9}{cm}^{-2}{s}^{-1}]\hfill &
4.53\hfill &4.79\hfill &5.15\hfill &4.20\cr
{\phi }_{\nu }(^{8}{B})~[{10}^{9}{cm}^{-2}{s}^{-1}]\hfill &
4.85\hfill &5.07\hfill &6.62\hfill &2.87\cr
{\phi }_{\nu }(^{13}{N})~[{10}^{9}{cm}^{-2}{s}^{-1}]\hfill &
4.07\hfill &2.50\hfill &6.18\hfill &3.94\cr
{\phi }_{\nu }(^{15}{O})~[{10}^{9}{cm}^{-2}{s}^{-1}]\hfill &
 3.45\hfill &3.38\hfill &5.45\hfill &3.88\cr
{\phi }_{\nu }(^{17}{F})~[{10}^{9}{cm}^{-2}{s}^{-1}]\hfill &
4.02\hfill &4.06\hfill &6.48\hfill &4.71\cr
\Sigma (\phi \sigma)_{Cl}~[SNU]\hfill
&\rm 7\pm 1\hfill &\rm 7\pm 1 \hfill&\rm 9.3\pm 1.4\hfill&
\rm 4.6\pm 1.6\hfill \cr
\Sigma(\phi\sigma)_{Ga}~[SNU]\hfill &\rm 127\pm 6\hfill
&\rm 128\pm 7\hfill &\rm 137\pm 8\hfill
&\rm 119\pm 7\hfill \cr}$$


\end{document}